# DISSERTATION

## **“Retinotopic Mechanics Derived Using Classical Physics”**

## ***“Retinotopische Mechanik, hergeleitet aus der klassischen Physik”***

zur Erlangung des akademischen Grades
**Doctor rerum medicinalium (Dr. rer. medic.)**

vorgelegt der Medizinischen Fakultät
Charité – Universitätsmedizin Berlin

von

**Ifedayo-EmmanuEL Adeyefa-Olasupo**

Betreuer: Prof. Dr. Henrik Walter

Datum der Promotion: ……………………

## Abstract

The concept of a cell's receptive field is a bedrock in systems neuroscience, and the classical static description of the receptive field has had enormous success in explaining the fundamental mechanisms underlying visual processing. Borne out by the spatio-temporal dynamics of visual sensitivity to probe stimuli in primates, I build upon this static account with the introduction of a new computational field of research, retinotopic mechanics. At its core, retinotopic mechanics assumes that, during active sensing, receptive fields are not static but can shift beyond their classical extent. Specifically, the canonical computations and the neural architecture that support these computations are inherently mediated by a neurobiologically inspired force field. For example, when the retina is displaced because of a saccadic eye movement from one point in space to another, cells across retinotopic brain areas are tasked with discounting the retinal disruptions such active surveillance inherently introduces. This neural phenomenon is known as spatial constancy. Using retinotopic mechanics, I propose that to achieve spatial constancy, or any active visually mediated task, retinotopic cells, namely their receptive fields, are constrained by eccentricity-dependent elastic fields. I propose that elastic fields are self-generated by the visual system and allow receptive fields the ability to predictively shift beyond their classical extent to future post-saccadic locations, such that neural sensitivity which would otherwise support intermediate eccentric locations likely to contain retinal disruptions is transiently blunted.

# 1. Background

## 1.1. Predictive Neural Remapping

The concept of a brain cell or neuron's classical receptive field (cRF) is fundamental in systems neuroscience and has been immensely successful in explaining the basic mechanisms of perception. In fact, seven Nobel Prizes have been awarded to researchers studying cRF structure and function: Sherrington, Ramón y Cajal, Granit, Hartline, von Békésy, Hubel, and Wiesel [1]. However, a seminal paper published by Michael Goldberg and colleagues [2] challenged this classical static description of the cRF. In this study, the authors demonstrated that prior to a saccadic eye movement (SEM), cRFs can transiently shift their neural sensitivity toward their future post-saccadic location, almost as if they had shifted their spatial preference. This phenomenon is known as retinotopic remapping, or what I will refer to here as predictive neural remapping (PNR). Its translational form has been proposed as a key mechanism that allows biological systems to maintain a stable and continuous perception of the visual world, despite the retinal displacement and motion blur that accompany each SEM. While this phenomenon has been replicated by other researchers, a study conducted by Tirin Moore and colleagues reported that cRFs primarily shift their sensitivity toward spatial extents surrounding the saccade target, not their future post-saccadic locations. This form of PNR is commonly referred to as convergent remapping [3].

Translational and convergent accounts of PNR remain at odds, and the functional role that these divergent forms of remapping play in mediating spatial constancy remains unresolved. Several critical questions persist. For instance, the high frequency of saccades (occurring approximately 2–4 times per second) and the associated cRF shifts may impose significant energetic demands. Yet the fundamental principles governing how the brain balances energy expenditure with the need for predictive accuracy remain unclear. Additionally, the functional architecture that allows retinotopic cells to maintain or enhance sensitivity beyond their classical centre-surround structure, while avoiding maladaptive forms of PNR, remains a mystery.

## 1.2. Mathematical Overview

In the following sections, I provide a concise overview of the most significant mathematical models related to PNR. Specifically, I review prominent studies that exemplify these models and reference additional studies employing similar mathematical approaches, although different in their formulations. After discussing these models, I briefly address their limitations. In Chapter 2, I introduce my mathematical framework, retinotopic mechanics (RM), and its advantages.

### 1.2.1. Vector Subtraction Models

The first class of mathematical models employed by theorists to investigate PNR can be classified as vector subtraction models (VSMs). These models suggest that PNR involves the manipulation of mathematical vectors through subtractive operations. Specifically, retinotopic cells, whether encoding the upcoming SEM [4], efference copy signals [5,6], or velocity signals required for the eyes to reach the saccade target [7], must have information that corresponds to the difference in magnitude between the saccade target and the current position of the eye. Using this information, cells across retinotopic brain areas (RBAs) can correctly predict and shift their neural response to their future post-saccadic locations.

Let $f(x_\varphi, y_\varphi, t)$ denote the response of a retinotopic cell at time t. If the cell correctly predicts the impending spatial extent of the saccade target, resulting in $f(x_\varphi + \delta x, y_\varphi + \delta y, t + \delta t)$ at a future time $t + \delta t$, then its response is invariant:

$$f(x_\varphi + \delta x,\ y_\varphi + \delta y,\ t + \delta t) = f(x_\varphi,\ y_\varphi,\ t) \qquad \text{[Eq. 1]}$$

Expanding on [Eq. 1], we can explicitly approximate this neural invariance over some future interval δt with respect to the displacement of a saccade target along the x and y directions. Specifically, δx and δy denote the target displacement at the current time point t, where $\partial f/\partial x$ and $\partial f/\partial y$ represent the corresponding response gradients with respect to x and y [Eq. 2]. Consistent with [Eq. 1], we can now directly observe that under certain conditions, for example small target displacements, the change in the response of the cell is negligible, consistent with the translational form of PNR.

$$f(x_\varphi, y_\varphi, t + \delta t) = f(x_\varphi, y_\varphi, t) - (\delta x)\frac{\partial f}{\partial x} - (\delta y)\frac{\partial f}{\partial y} \qquad \text{[Eq. 2]}$$

Let us further assume that the saccade target is fixed (that is, no saccade displacement signal is present) and that f is driven by signals carrying information about velocity, so that $\delta x = -\delta t \cdot v_x$ and $\delta y = -\delta t \cdot v_y$. With this assumption it is possible to modify [Eq. 2], where in [Eq. 3] the change over some time interval δt is the result of velocity signals ($V_x$, $V_y$) coming from relevant RBAs. To approximate the discrete dynamics for a given cell i, we compute its future response as a weighted sum of the activity of its neighbouring cells j. Here, the coefficients $a_{ij}$, $b_{ij}$ and $c_{ij}$ are weights that depend on the distance between i and j [Eq. 4].

$$f(x_\varphi, y_\varphi, t + \delta t) = f(x_\varphi, y_\varphi, t) + \delta t\left(V_x\frac{\partial f}{\partial x} + V_y\frac{\partial f}{\partial y}\right) \qquad \text{[Eq. 3]}$$

$$f(x_i, y_i, t + \delta t) = \sum_j \left(a_{ij} + \delta t\, V_x b_{ij} + \delta t\, V_y c_{ij}\right) f\left(x_i + dx_j,\ y_i + dy_j,\ t\right) \quad \text{[Eq. 4]}$$

### 1.2.2. Artificial Network Models

The second class of models that has been used to investigate PNR comprises artificial network models (ANMs). These models typically include a recurrent hidden layer along with an input and output layer [8,9,10]. The input layer in general encodes information about the initial saccade target location. Thus, the neural activity at any spatial extent (x, y) in the retinotopic field is represented by f(x, y), where $x_0$ and $y_0$ correspond to the cells most strongly activated by the saccade target, while σ specifies the width of the response distribution. This central activity represents the area of the retina directly "viewing" the saccade target. Adjacent cells, which are less activated by the target, contribute to a gradient response profile. The spread of this response is modelled by a Gaussian:

$$f(x, y) = \exp\left(-\frac{(x - x_0)^2 + (y - y_0)^2}{2\sigma^2}\right) \quad \text{[Eq. 5]}$$

Efference copy information about the saccade target is fed into the hidden layer, where it is encoded in one of three ways: as a transient signal about the impending SEM, a sustained signal that plans the SEM from the centre of gaze to the saccade target, or a velocity signal. Let $h_j(t+1)$ represent the response of retinotopic cell $h_j$ at time t+1. The recurrent weights between cells within the hidden layer are denoted by $w_{ij}$, while $\varepsilon_k(t)$ and $w'_{jk}$ represent the efference-copy signals from the input layer and their corresponding weights, respectively. With this architecture, cells in the hidden layer can adjust the internal representation that approximates the location of the saccade target, $x_0$, $y_0$. This adjustment leads to an updated internal representation that accurately reflects the spatial extent of the target once the eye has landed.

$$h_j(t + 1) = \phi\left(\sum_i w_{ij}\, h_i(t) + \sum_k w'_{jk}\, \varepsilon_k(t)\right) \quad \text{[Eq. 6]}$$

The updated internal representation $x_0'$, $y_0'$ is made available to units within the output layer. Here, δx and δy represent the impending displacement of the eyes. To assess whether the artificial network has accurately remapped to the target location, the updated internal representation is evaluated and $r_\varepsilon$, the remapping error, is computed. Here, ω denotes the predicted representation and $\hat{\omega}$ the correct representation. Because the aim of these models is to minimise $r_\varepsilon$ through iterative adjustments (i.e. learning), the partial derivative of $r_\varepsilon$ with respect to each weight in all layers (input, hidden and output) is computed. Over time, weights across all layers are updated using gradient descent, where $w'_{ij}$ represents the updated weights and η denotes the learning rate used by the artificial network.

$$x_0^{'} = x_0 + \delta x, \qquad y_0^{'} = y_0 + \delta y \qquad \text{[Eq. 7]}$$

$$r_\varepsilon = \sqrt{\frac{1}{n}\sum_{j=1}^{n}\left(\omega_j - \widehat{\omega}_j\right)^2} \qquad \text{[Eq. 8]}$$

$$\frac{\partial r_\varepsilon}{\partial w_{ij}} = \frac{\partial r_\varepsilon}{\partial \omega_j} \cdot \frac{\partial \omega_j}{\partial w_{ij}} \qquad \text{[Eq. 9]}$$

$$w_{ij}^{'} = w_{ij} - \eta \frac{\partial r_\varepsilon}{\partial w_{ij}} \qquad \text{[Eq. 10]}$$

### 1.2.3. Biologically Inspired Models

The third and final class of models typically used to investigate PNR I will refer to as biologically inspired models (BIMs). These models are biologically inspired because they are constrained by findings from psychophysical and neurophysiological studies in primates [11,12,13]. Specifically, they assume that RBAs have a shared retinotopic map, reflecting the spatial layout of the retina. They further assume that visual processing is hierarchical, with lower-order RBAs (α) responsible for processing finer details through smaller cRFs, while higher-order RBAs (β), with larger cRFs, pool information from lower-order areas to generalise across broader regions of the visual field. Each RBA is characterised by a magnification factor, reflecting a higher concentration of sensitised cells within the central 2 degrees of the visual field compared with the periphery. Let $r_i^{\alpha 1}$ represent the response of retinotopic cells, where $S_p$ denotes the position of a saccade target with magnitude k. The term $c_i^{\alpha 1}$ indicates the centre of a cell's cRF, whose density scales with eccentricity. Indeed, the activity of cell i decreases as the distance between the saccade target and $cRF_i$ increases [Eq. 11].

$$r_i^{\alpha_1} = k\exp\left(\frac{-\left\|S_p - c_i^{\alpha_1}\right\|^2}{2\left(\sigma_i^{\alpha_1}\right)^2}\right) \qquad \text{[Eq. 11]}$$

As a new region of interest is selected, oculomotor feedback signals, the precursors to gain modulation signals [Eq. 12], provide crucial information about the consequence of the impending SEM. In this context, $c^{SEM}$ denotes the SEM signal: cells closer to the saccade target receive a stronger feedback signal, and those farther away a weaker signal. The feedback signal, modulated by f(t), strengthens as the time to saccade onset approaches and declines shortly after.

$$\hat{r}_i^{\alpha} = \exp\left(\frac{-\left\|c_i^{\alpha_1} - c^{\text{SEM}}\right\|^2}{2\left(\sigma_i^{\alpha_1}\right)^2}\right) f(t) \qquad \text{[Eq. 12]}$$

Once feedback signals emerge, a gain modulation is applied to the visual input $r_i^{\alpha 1}$, leading to an amplified response around the saccade target:

$$r_i^{\lambda_1} = \frac{r_i^{\alpha_1}\left(1 + k\,\hat{r}_i^{\alpha}\right)}{1 + k \max_j\left(r_j^{\alpha_1}\right)\hat{r}_i^{\alpha}} \qquad \text{[Eq. 13]}$$

The gain-modulated activity of cells in lower-order RBAs is eventually pooled, providing a spatially invariant representation in higher-order RBAs. Here, $r_j^{\varphi}$ represents the pooled activity, which is dominated by the cell i with the largest gain-modulated response $r_i^{\lambda 1}$.

$$r_j^{\varphi} = \max_i\left(r_i^{\lambda_1} \exp\left(-\frac{\left\|c_i^{\alpha_1} - c_j^{\varphi_1}\right\|^2}{2\left(\sigma_j^{\varphi_1}\right)^2}\right)\right) \qquad \text{[Eq. 14]}$$

## 1.3. Model Limitations

Despite the computational insights these three classes of models have provided, they exhibit four significant limitations. First, these models are highly parameterised and include non-modular components, with simulations ultimately conforming to specific experimental demands and thereby favouring one form of PNR over another. For example, VSMs and ANMs cannot account for the convergent form of PNR, while BIMs fail to explain the translational form. As a result, none of these models adequately account for the divergence in experimental findings. Second, the frequency of saccades and the accompanying predictive shifts in cRFs are energetically costly, yet these models do not identify a governing principle by which retinotopic cells balance energy expenditure against predictive function. Third, while the anatomical centre-surround structure of cRFs is well established, none of these models propose a functional neural architecture that can explain how a cRF shifts beyond its anatomical extent without leading to unsustainable forms of PNR. Finally, these models are neither general nor expressive enough, as they cannot account for afoveated biological systems, which also execute SEMs in order to orient themselves in the world.

# 2. Method

In this chapter, I address the limitations of the models discussed in the previous chapter by introducing a fundamentally different mathematical framework, retinotopic mechanics. This framework is rooted in the principles and formalism of theoretical physics. Building upon this foundation, I construct a general model in Chapter 3 that incorporates the new framework. Additionally, I compare the model's predictions with both existing experimental data and data from psychophysical experiments that I have personally conducted.

## 2.1. Nonlinear Neural Spring Dynamics

Retinotopic mechanics assumes that retinotopic cells routinely shift beyond their classical centre-surround structure whenever the animal is moving, which includes cases when it is moving its eyes. A retinotopic cell does not behave like a simple thermometer, driven solely by the absence or presence of external perturbation, nor does it behave like a perceptron. Instead, this framework posits that retinotopic cells behave computationally like a spring-loaded sensor that is governed by neural elastic fields (εlφ's).

### 2.1.1. Adaptive Elasticity

In the case where a cell's εlφ is adaptive, let $\hat{x}$ represent the displacement of the neural sensor within its εlφ, and k denote its spring constant. As $\hat{x}$ approaches the boundary of its εlφ, the sensor paradoxically ceases to respond despite the presence of external forces. Such inhibition by the sensor's εlφ prevents unsustainable forms of PNR, such as over-translational shifts, and ensures that cells across RBAs remain appropriately sensitive across the entire visual field.

$$\mathbf{F}_S = \begin{cases} -k\hat{x} & \text{if } |\hat{x}| \leq \varepsilon l\varphi \\ 0 & \text{if } |\hat{x}| > \varepsilon l\varphi \end{cases} \qquad \text{[Eq. 15]}$$

### 2.1.2. Maladaptive Elasticity

Conversely, a cell's εlφ can also become maladaptive in one of two ways, reflecting dynamics seen in patients with mental and psychiatric disorders [14,15]. In the first case, let τ represent the spring's delay in response: the spring responds to an external force only after a latency τ has elapsed, and remains unresponsive before that point (Fig. 1B).

$$\mathbf{F}_S = \begin{cases} -k\hat{x}(t-\tau) & \text{if } |\hat{x}(t-\tau)| \leq \varepsilon l\varphi \text{ and } t > \tau \\ 0 & \text{if } |\hat{x}(t-\tau)| > \varepsilon l\varphi \text{ or } t \leq \tau \end{cases} \qquad \text{[Eq. 16]}$$

The second way a spring's εlφ can become maladaptive is when the spring is displaced beyond its εlφ. As the spring approaches its boundaries, it enters a non-linear regime, where the additional quadratic term $-\gamma(\hat{x} - \varepsilon l\varphi)^2$ starts to dominate, causing the spring to move beyond its εlφ. In theory, this would require an increasingly large force, but in a biological or maladaptive context the force exerting its influence on the spring is hallucinated, akin to how a clinical patient might hallucinate the presence of an external cue when, in reality, no such external perturbation exists.

$$\mathbf{F}_S = \begin{cases} -k\hat{x} & \text{if } |\hat{x}| \leq \varepsilon l\varphi \\ -k\hat{x} - \gamma(\hat{x} - \varepsilon l\varphi)^2 & \text{if } |\hat{x}| > \varepsilon l\varphi \end{cases} \quad \text{[Eq. 17]}$$

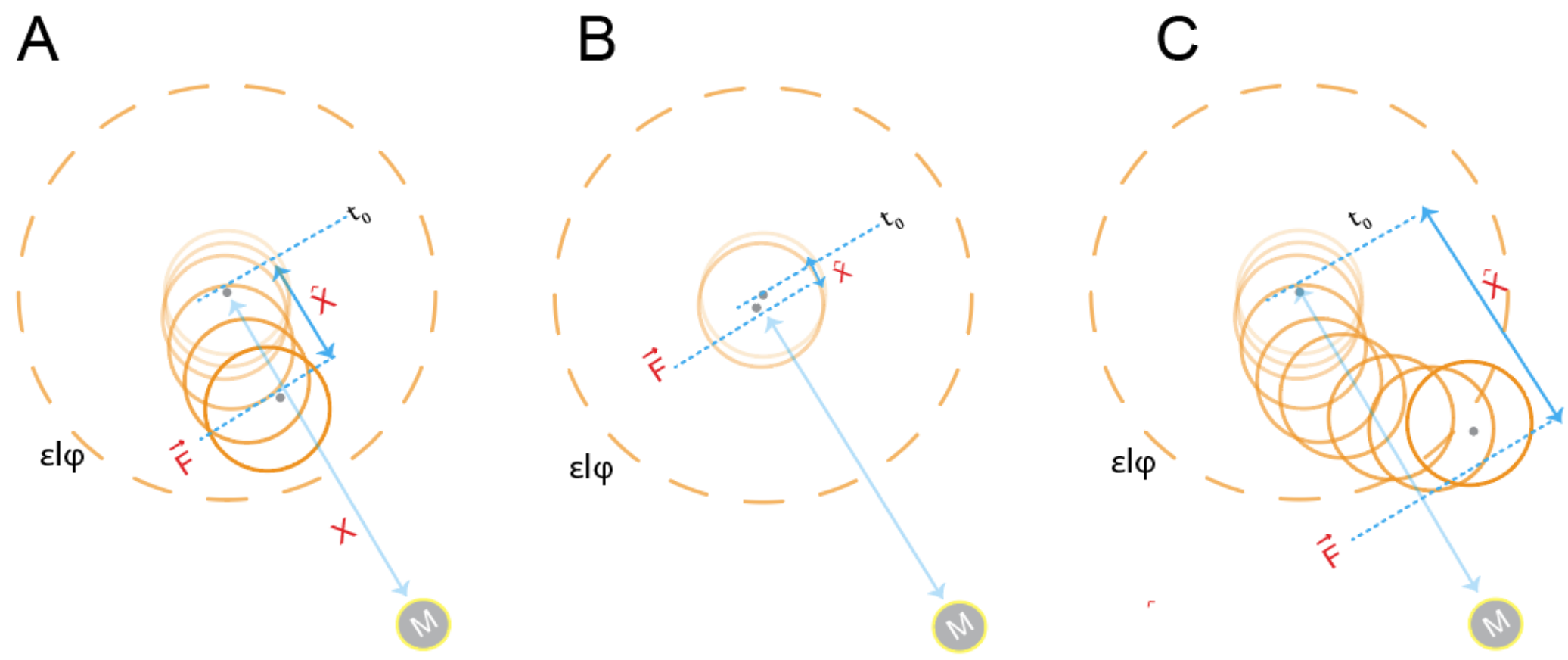

*Figure 1. Non-linear dynamics of spring-loaded neural sensors (own representation, Ifedayo-Emmanuel Adeyefa-Olasupo).*

## 2.2. Differences in Design Principles

While foveated and afoveated biological systems execute SEMs in order to orient themselves within their environment, they have evolved with different design principles. These principles determine how retinotopic cells, and in turn their εlφ's, tile visual space in the absence of external forces [16].

Let us assume that each cRF across RBAs actively modulates sensitivity across the visual field and is defined by two key parameters. The first, denoted $p_i$, represents the centre location $(x_i, y_i)$ of the cRF at time $t_0$. The location $p_i$ directly determines the cRF's eccentricity, represented as $ecc_i$. The second parameter is $r_i$, the radius of the cRF. In foveated biological systems $r_i$ is a function of eccentricity $ecc_i$, where k represents the rate at which $r_i$ scales with increasing eccentricity [Eq. 18]. In more dynamic cases, such as around the time of a SEM, $r_i$ can vary over time, influenced by the rate of change of $ecc_i$ [Eq. 19].

$$\frac{dr_i}{d(\mathrm{ecc}_i)} = k \qquad \text{[Eq. 18]}$$

$$\frac{dr_i}{dt} = k\,\frac{d(\mathrm{ecc}_i)}{dt} \qquad \text{[Eq. 19]}$$

As described in Section 2.1.1, each cRF possesses an $\varepsilon l\varphi$. Throughout what follows, $\varepsilon l\varphi$ denotes the radius of the outer functional boundary of that elastic field, so that it is directly comparable with $r_i$. In biological systems with a fovea, $\varepsilon l\varphi$ is dependent on $\mathrm{ecc}_i$. In this context, $r_i$ is scaled globally by a factor $\Psi$, ensuring that the radius of the $\varepsilon l\varphi$ exceeds that of the cRF, where $\Psi > 1$. This scaling ensures that the $\varepsilon l\varphi$ expands appropriately with increasing $\mathrm{ecc}_i$. In more dynamic scenarios, the rate of change of $\varepsilon l\varphi$ over time is directly related to the rate of change of $\mathrm{ecc}_i$, with $\Psi$ scaling this relationship, as shown in [Eq. 21].

$$\varepsilon l\varphi_i = \Psi\, r_i, \qquad \Psi > 1 \qquad \text{[Eq. 20]}$$

$$\frac{d(\varepsilon l\varphi_i)}{dt} = \Psi\,\frac{dr_i}{dt} = \Psi\, k\,\frac{d(\mathrm{ecc}_i)}{dt} \qquad \text{[Eq. 21]}$$

In contrast, in afoveated biological systems the relationship is independent of $\mathrm{ecc}_i$, meaning that $r_i$ remains constant regardless of variations in $\mathrm{ecc}_i$. Mathematically, this can be expressed as follows:

$$\frac{dr_i}{d(\mathrm{ecc}_i)} = 0 \qquad \text{[Eq. 22]}$$

$$\frac{dr_i}{dt} = 0 \qquad \text{[Eq. 23]}$$

In such systems, similar to the foveated system, the $\varepsilon l\varphi$ can still be described by [Eq. 20]. However, since $r_i$ is constant, $\varepsilon l\varphi$ also remains unchanged. Therefore, the rate of change of $\varepsilon l\varphi$ over time is given by:

$$\frac{d(\varepsilon l\varphi_i)}{dt} = 0 \qquad \text{[Eq. 24]}$$

## 2.3. Retinotopic Mass

For cRFs to operate effectively within a force field such that they can move beyond their classical centre-surround structure, the presence of what I term retinotopic mass (rm) is crucial. In this context, rm exhibits three distinct phases: growth, plateau and decay (Fig. 2, right panel). Let g represent the rate at which the mass appears, whether suddenly or gradually, and let c be a scaling factor that modulates the overall magnitude of the growth rate. During the growth phase, the rate of change of retinotopic mass is described by:

$$\frac{d(rm)}{dt} = g\,c\,t^{g-1} \qquad \text{[Eq. 25]}$$

The plateau phase, in which rm remains constant, is represented by:

$$\frac{d(rm)}{dt} = 0 \qquad \text{[Eq. 26]}$$

Finally, the decay phase is characterised by a negative rate, where λ represents the rate of mass decay and c continues to modulate the magnitude of the decay. This is expressed as:

$$\frac{d(rm)}{dt} = -\lambda\,c\,t^{-(\lambda+1)} \qquad \text{[Eq. 27]}$$

## 2.4. Neural Force Field

The presence of retinotopic masses gives rise to a force field that displaces cRFs from their equilibrium positions. In theory, cRFs can be influenced by an infinite number of retinotopic masses, either concurrently or successively, each contributing to the overall perturbation. For instance, a single rm located near the target of an impending saccadic eye movement can be used to study the convergent form of PNR. Additionally, a virtual rm positioned at infinity, aligned with the SEM direction, may be considered to study its impact in a manner consistent with the translational form of PNR. These configurations can also coexist, enabling the exploration of their interactions within overlapping temporal windows. Such simulation examples underscore the expressiveness and modularity of retinotopic mechanics as a framework, ensuring that different types of PNR can be independently explored and that the model does not inherently favour any specific form of PNR, a limitation of the frameworks described in Chapter 1.

# 3. General Model

In Chapter 1, I provided an overview of three mathematical frameworks and the accompanying models that implement them, which researchers have used to investigate PNR. In Chapter 2, I introduced the mathematical framework of retinotopic mechanics, highlighting its core advantages. Here, I construct a general model of PNR that implements retinotopic mechanics. This model not only resolves divergent observations in the literature, such as the focus on translational and convergent accounts, but also uncovers the fundamental principles governing how the brain balances energy expenditure with the need for predictive accuracy. Finally, this general model clarifies the role εlφ's play in ensuring that cRFs are appropriately sensitised across visual space.

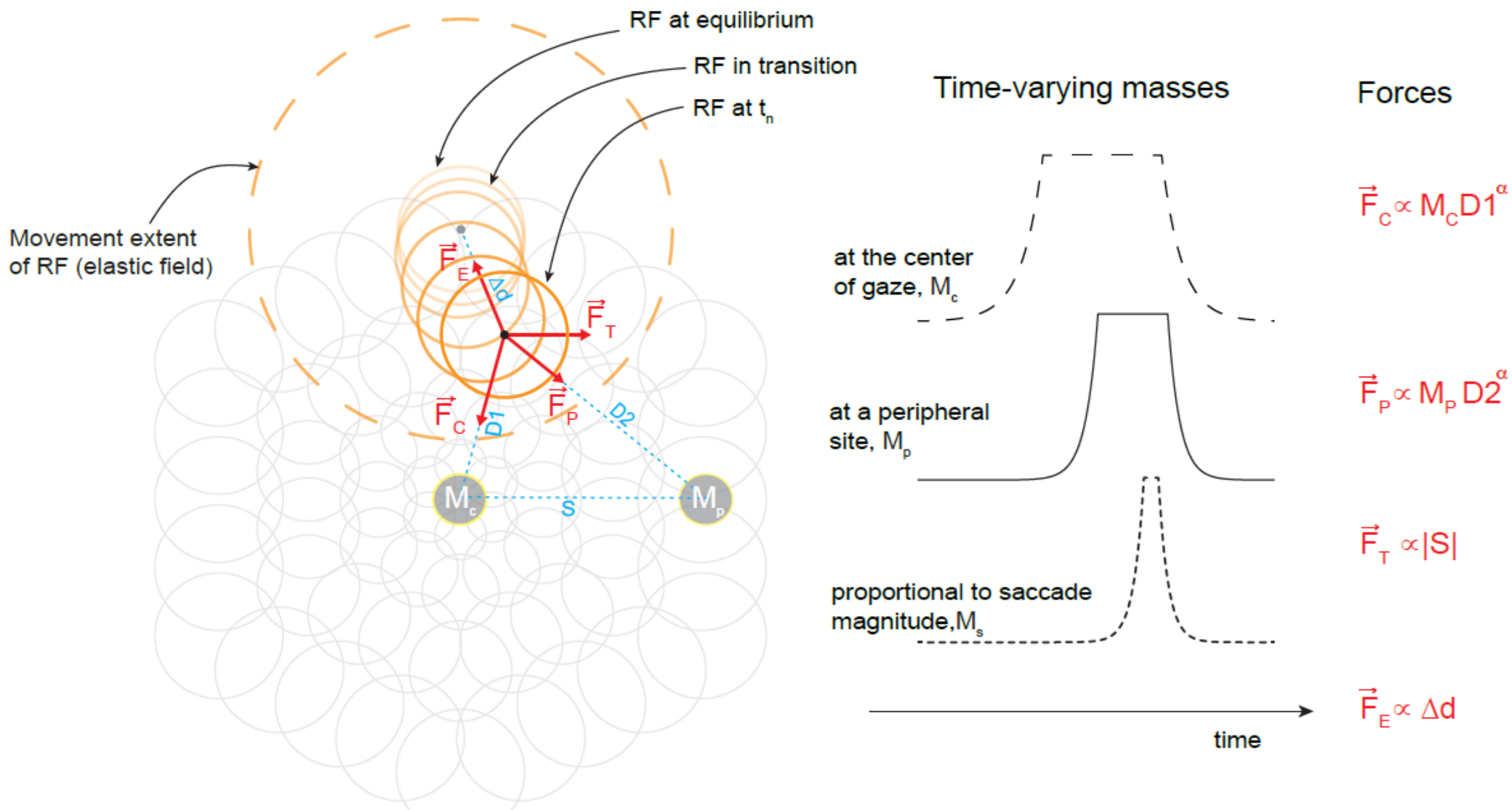

*Figure 2. General model of remapping using retinotopic mechanics (own representation, Ifedayo-Emmanuel Adeyefa-Olasupo).*

## 3.1. Core Assumptions

The general model includes three forces (centripetal, convergent and translational) that influence the population of cRFs within distinct but temporally overlapping windows. First, we assume that the biological system architecture being modelled is a foveated system, in which the radius $r_i$ of each cRF scales with eccentricity ($ecc_i$). Second, we assume an adaptive εlφ that also scales with eccentricity. Third, we assume that the centripetal force is the first to exert its influence on the cRFs, shifting them toward the current centre of gaze, followed by the convergent force, which moves the cRFs toward the saccade target, and finally by the translational force, which shifts the cRFs in the direction of the impending saccade. In the context of PNR this means that PNR includes centripetal, convergent and translational components. Finally, we assume that the neural force field obeys an inverse-distance law, $1/r^{1.6}$, akin to Newton's law of universal gravitation. This law governs the two forces whose retinotopic masses occupy finite positions in visual space, namely the centripetal and convergent forces. The translational force is the exception: because its mass lies at infinity there is no finite separation to invert, and it instead scales linearly with the magnitude of the impending saccade [Eq. 38]. Throughout Chapter 3 the inverse law is implemented by the distance exponent α, which takes the value $\alpha = -1.6$, so that $r^{\alpha} = 1/r^{1.6}$ and the magnitude of each of these two forces decreases with increasing distance from its retinotopic mass.

### 3.1.1. Centripetal Force

Let $\mathbf{f}_C$ represent the centripetal force that displaces $cRF_i$ from its equilibrium position toward the central 2° of the visual field. This force is defined in terms of its magnitude and direction, with the expression given by:

$$\mathbf{f}_C = k_{c\beta}\, \mathbf{U}_{\mathrm{rm}C}\, rm_C\, r_1^{\alpha} \quad \text{[Eq. 28]}$$

Here, $k_{c\beta}$ denotes the magnitude of the centripetal force and $\mathbf{U}_{\mathrm{rmC}}$ is the unit vector in the direction of $r_1$, the spatial displacement of $cRF_i$ from $rm_C$, raised to the distance exponent $\alpha = -1.6$. The differential form of the centripetal force is derived by taking the partial derivative with respect to $r_1$:

$$d\mathbf{f}_C = \frac{\partial \mathbf{f}_C}{\partial r_1}\, dr_1 \quad \text{[Eq. 29]}$$

This expression represents the change in the centripetal force as a function of the spatial displacement $r_1$. The partial derivative of the centripetal force is given by:

$$\frac{\partial \mathbf{f}_C}{\partial r_1} = k_{c\beta}\, rm_C\, \frac{\partial}{\partial r_1}\left(\mathbf{U}_{\mathrm{rm}C}\, r_1^{\alpha}\right) \quad \text{[Eq. 30]}$$

Applying the power rule to the term $r_1^{\alpha}$ results in:

$$\frac{\partial}{\partial r_1}\left(r_1^{\alpha}\right) = \alpha\, r_1^{\alpha-1} \quad \text{[Eq. 31]}$$

Substituting this into the expression for the centripetal force yields the differential form:

$$d\mathbf{f}_C = k_{c\beta}\, rm_C\, \alpha\, \mathbf{U}_{\mathrm{rm}C}\, r_1^{\alpha-1}\, dr_1 \quad \text{[Eq. 32]}$$

This formulation captures the dependence of the differential centripetal force on the spatial displacement $r_1$ and the exponent α, reflecting the dynamics that govern the behaviour of cRFs under centripetal forces.

### 3.1.2. Convergent Force

The centripetal force is followed by the convergent force, where $k_{p\beta}$ represents the convergent constant. Let $\mathbf{U}_{\mathrm{rm}P}$ denote the unit vector in the direction of $r_2$, the spatial displacement of $cRF_i$ from $rm_P$, raised to α. The convergent force can be expressed as:

$$\mathbf{f}_P = k_{p\beta}\, \mathbf{U}_{\mathrm{rm}P}\, rm_P\, r_2^{\alpha} \qquad \text{[Eq. 33]}$$

Here, the differential form of $\mathbf{f}_P$ is derived by taking the partial derivative with respect to $r_2$:

$$d\mathbf{f}_P = \frac{\partial \mathbf{f}_P}{\partial r_2}\, dr_2 \qquad \text{[Eq. 34]}$$

Analogous to the derivation in the centripetal case, this equation describes the change in the convergent force as a function of $r_2$. The partial derivative of the convergent force is given by:

$$\frac{\partial \mathbf{f}_P}{\partial r_2} = k_{p\beta}\, rm_P\, \frac{\partial}{\partial r_2}\left(\mathbf{U}_{\mathrm{rm}P}\, r_2^{\alpha}\right) \qquad \text{[Eq. 35]}$$

Applying the power rule for differentiation yields:

$$\frac{\partial}{\partial r_2}\left(r_2^{\alpha}\right) = \alpha\, r_2^{\alpha-1} \qquad \text{[Eq. 36]}$$

Substituting this result back into the expression, the differential form becomes:

$$d\mathbf{f}_P = k_{p\beta}\, rm_P\, \alpha\, \mathbf{U}_{\mathrm{rm}P}\, r_2^{\alpha-1}\, dr_2 \qquad \text{[Eq. 37]}$$

This formulation highlights the sensitivity of the differential force to the spatial displacement $r_2$ and α. Furthermore, it captures the progressive attenuation of the force as $r_2$ increases, so that cRFs lying farther from $rm_P$ are perturbed less strongly.

### 3.1.3. Translational Force

Unlike $\mathbf{f}_C$ and $\mathbf{f}_P$, the translational force $\mathbf{f}_T$ exerts its influence due to a mass at infinity rather than at an exact spatial extent in visual space. Specifically, the translational force can be expressed as:

$$\mathbf{f}_T = k_{T\beta}\, \mathbf{U}_{\mathrm{rm}T}\, rm_T |s| \qquad \text{[Eq. 38]}$$

Here, $k_{T\beta}$ represents the strength of this force, $\mathbf{U}_{\mathrm{rm}T}$ is the unit vector in the direction of $rm_T$, and |s| denotes the magnitude of the impending SEM. The differential form of $\mathbf{f}_T$ is obtained by taking the partial derivative with respect to |s|:

$$d\mathbf{f}_T = \frac{\partial \mathbf{f}_T}{\partial |s|}\, d|s| \quad \text{[Eq. 39]}$$

This equation describes the infinitesimal change in the translational force as a function of |s|. Taking the partial derivative yields:

$$\frac{\partial \mathbf{f}_T}{\partial |s|} = k_{T\beta}\, rm_T \frac{\partial}{\partial |s|}(\mathbf{U}_{\mathrm{rm}T}|s|) \quad \text{[Eq. 40]}$$

Since $\mathbf{U}_{\mathrm{rmT}}$ is a constant unit vector, the derivative simplifies to:

$$\frac{\partial}{\partial |s|}(|s|) = 1 \quad \text{[Eq. 41]}$$

Replacing this result into the expression for the differential form, we obtain:

$$d\mathbf{f}_T = k_{T\beta}\, rm_T\, \mathbf{U}_{\mathrm{rm}T}\, d|s| \quad \text{[Eq. 42]}$$

This formulation shows that the translational force varies linearly with changes in |s|. It allows the translational force to govern the motion of cRFs along the saccade direction, with the extent of the resulting shift bounded by εlφ through the spring force.

### 3.1.4. Spring Force (Internal Dynamics)

The spring force $\mathbf{F}_S$ governs the internal dynamics of $cRF_i$, in contrast to the external forces that perturb it from its equilibrium extent. In its simplest form, this internal dynamic is defined as:

$$\mathbf{F}_S = \begin{cases} -k_s\hat{x} & \text{for } |\hat{x}| \le \varepsilon l\varphi \\ 0 & \text{for } |\hat{x}| > \varepsilon l\varphi \end{cases} \quad \text{[Eq. 43]}$$

To express the differential form of the spring force, we first compute its partial derivative with respect to the displacement x̂:

$$d\mathbf{F}_S = \frac{\partial \mathbf{F}_S}{\partial \hat{x}}\, d\hat{x} \quad \text{[Eq. 44]}$$

This derivative quantifies the incremental change in the spring force with a small variation in x̂. The partial derivative can be expressed as:

$$\frac{\partial \mathbf{F}_S}{\partial \hat{x}} = \begin{cases} -k_s & \text{for } |\hat{x}| \le \varepsilon l\varphi \\ 0 & \text{for } |\hat{x}| > \varepsilon l\varphi \end{cases} \quad \text{[Eq. 45]}$$

Substituting the spring force expression into the partial derivative, and focusing on the case where $|\hat{x}| \le \varepsilon l\varphi$:

$$\frac{\partial}{\partial \hat{x}}(-k_s \hat{x}) = -k_s \qquad \text{[Eq. 46]}$$

Thus, the differential form of the spring force can be expressed as:

$$d\mathbf{F}_S = -k_s\, d\hat{x}, \qquad |\hat{x}| \leq \varepsilon l\varphi \qquad \text{[Eq. 47]}$$

This shows that the spring force acts linearly within the defined displacement range, providing a restoring influence back toward the equilibrium extent of $cRF_i$. In the case where $|\hat{x}|$ exceeds $\varepsilon l\varphi$ the spring force is withdrawn, consistent with [Eq. 43] and with the adaptive elasticity described in Section 2.1.1: it is the boundary of the elastic field itself, and not a growing restoring force, that arrests further displacement and thereby prevents unsustainable forms of PNR.

### 3.1.5. Net Force

The net force, denoted $\mathbf{F}_N$, represents the cumulative effect of the various forces acting on $cRF_i$. Specifically, this force integrates contributions from the centripetal force $\mathbf{f}_C$, convergent force $\mathbf{f}_P$, translational force $\mathbf{f}_T$ and spring force $\mathbf{F}_S$, reflecting the overall influence on $cRF_i$. The net force can be expressed as:

$$\mathbf{F}_N = \sum_{k=1}^{n} \mathbf{F}_k \qquad \text{[Eq. 48]}$$

For the present model this sum comprises four terms: the three external forces, which act on $cRF_i$ without restriction, and the spring force, which confines the resulting displacement to within $\varepsilon l\varphi$:

$$\mathbf{F}_N = \mathbf{f}_C + \mathbf{f}_P + \mathbf{f}_T + \mathbf{F}_S \qquad \text{[Eq. 49]}$$

Finally, the differential form of the net force can be derived by differentiating each component force:

$$d\mathbf{F}_N = d\mathbf{f}_C + d\mathbf{f}_P + d\mathbf{f}_T + d\mathbf{F}_S \qquad \text{[Eq. 50]}$$

# 4. Results

## 4.1. Density (Sensitivity) Readout

A bivariate Gaussian function can serve as an effective model of the changes in neural sensitivity that follow the displacement of $cRF_i$ by the net force $\mathbf{F}_N$. This function G(u) captures the spatial distribution of sensitivity around $cRF_i$ and can represent the increase or decrease in sensitivity as a response to the displacement. The Gaussian function is given by:

$$G(u) = \frac{1}{2\pi} \exp\left(-\frac{1}{2} u^{\mathrm{T}} u\right) \quad \text{[Eq. 51]}$$

Here, 1/2π is the normalisation factor that ensures the total volume under the bivariate Gaussian is equal to 1. The term $-½\, u^{T}u$ measures the squared Euclidean distance of u from the centre of the Gaussian. As u moves further from the centre, G(u) decreases rapidly because the exponential function suppresses values far from the centre. Next, we modify this basic Gaussian by introducing a bandwidth B, which controls the spread of the Gaussian. Here, B is a symmetric positive-definite 2 × 2 bandwidth matrix, and |B| denotes its determinant. This gives the scaled Gaussian function:

$$G_B(u) = |B|^{-1/2} G\left(B^{-1/2} u\right) \quad \text{[Eq. 52]}$$

The factor $B^{-1/2}$ is applied to the input u, stretching or compressing the Gaussian according to the value of B. A larger bandwidth results in a wider Gaussian, meaning that the cRF has a broader influence over a larger area; a smaller bandwidth results in a narrower Gaussian, making the cRF's influence more localised. The outer factor $|B|^{-1/2}$ ensures that the total volume under the two-dimensional Gaussian remains equal to 1, preserving the properties of a probability density function. Finally, to estimate the overall neural density at a specific retinotopic location $r_l$, we sum the contributions of multiple neighbouring cRFs using the following equation:

$$\widehat{D}(r_l, B) = \frac{1}{n} \sum_{i=1}^{n} G_B\left(r_l - r_l^i\right) \quad \text{[Eq. 53]}$$

Here, each cRF's contribution to $r_l$ is determined by $G_B(r_l - r_l^i)$, where $r_l^i$ represents the centre of the i-th cRF. The function calculates how much the i-th cRF contributes to the density at $r_l$, decreasing as the distance between $r_l$ and $r_l^i$ increases. The scaling factor 1/n ensures that the final value, which is the sum of contributions from all $cRF_i$, is an average (population activity or density).

## 4.2. Putative Sensitivity

Changes in density (or sensitivity) at four distinct extents, presumed to map along the radial axis of visual space, were measured as $cRF_i$ were modulated by the influence of $\mathbf{F_N}$. In Figure 3, the brightest dot represents the foveal outer extent ($F_{out}$), located to the left of the centre of gaze. The second brightest dot represents the foveal inner extent ($F_{in}$). Next is the peripheral inner extent ($P_{in}$), positioned left of the saccade target, followed by the peripheral outer extent ($P_{out}$).

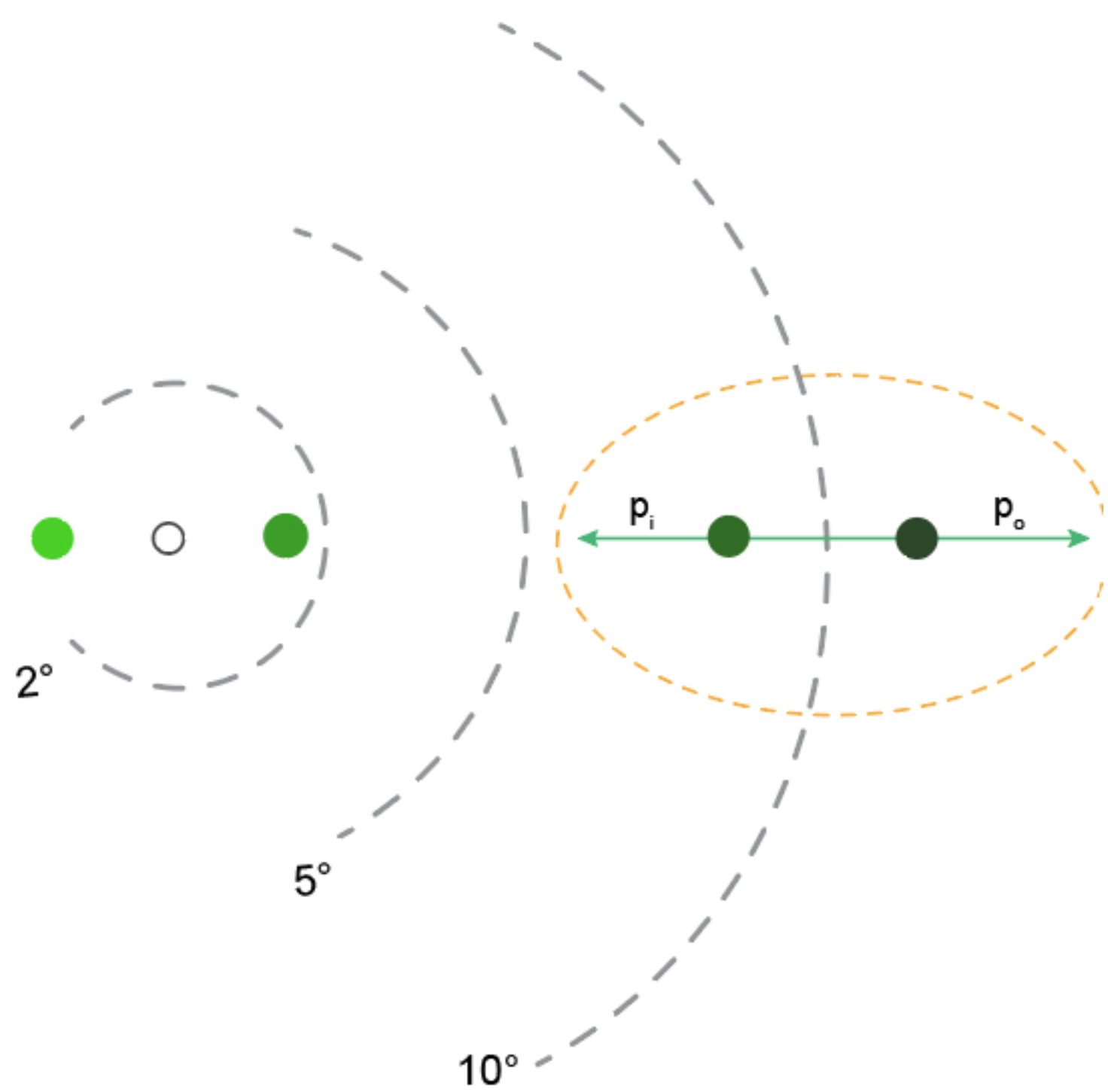

*Figure 3. Radial extents sampled (own representation, Ifedayo-Emmanuel Adeyefa-Olasupo).*

### 4.2.1. Density Predictions at Foveal Extents

When $cRF_i$ are subjected to a centripetal force between 150 and 270 ms before the force declines, density levels at $F_{in}$ and $F_{out}$ remain relatively stable, as expected. Following the centripetal force, a convergent force acts to shift $cRF_i$ toward 10 degrees along the radial axis of the visual field. Subsequently, with the activation of a translational force that further shifts $cRF_i$ toward the peripheral region, a rapid decline in density is observed at the foveal extent. Interestingly, $F_{out}$, being the farthest from the saccade target location, exhibits a slightly higher density level compared with $F_{in}$, which is closer to the periphery (i.e. the new centre of gaze).

### 4.2.2. Density Predictions at Peripheral Extents

Between 150 and 270 ms before the decline of the centripetal force, a noticeable yet modest decline in density levels at $P_{in}$ is observed. Around 240 ms, with the activation of the convergent force directing most, but not all, $cRF_i$ back toward the periphery, a slight rebound in density levels at this location occurs. This is followed by a gradual increase in density as the translational force begins to take effect. At the $P_{out}$ extent, the measured location that is furthest from the centripetal mass, a significant decline in density levels is observed. Around 240 ms, as the convergent force begins to influence $cRF_i$, a slight rebound in density levels occurs. Interestingly, as the translational force subsequently exerts its influence on $cRF_i$, this more peripheral extent, though expected to receive resource transfers later from the central region, exhibits a rapid and pronounced increase in density levels compared with the $P_{in}$ extent.

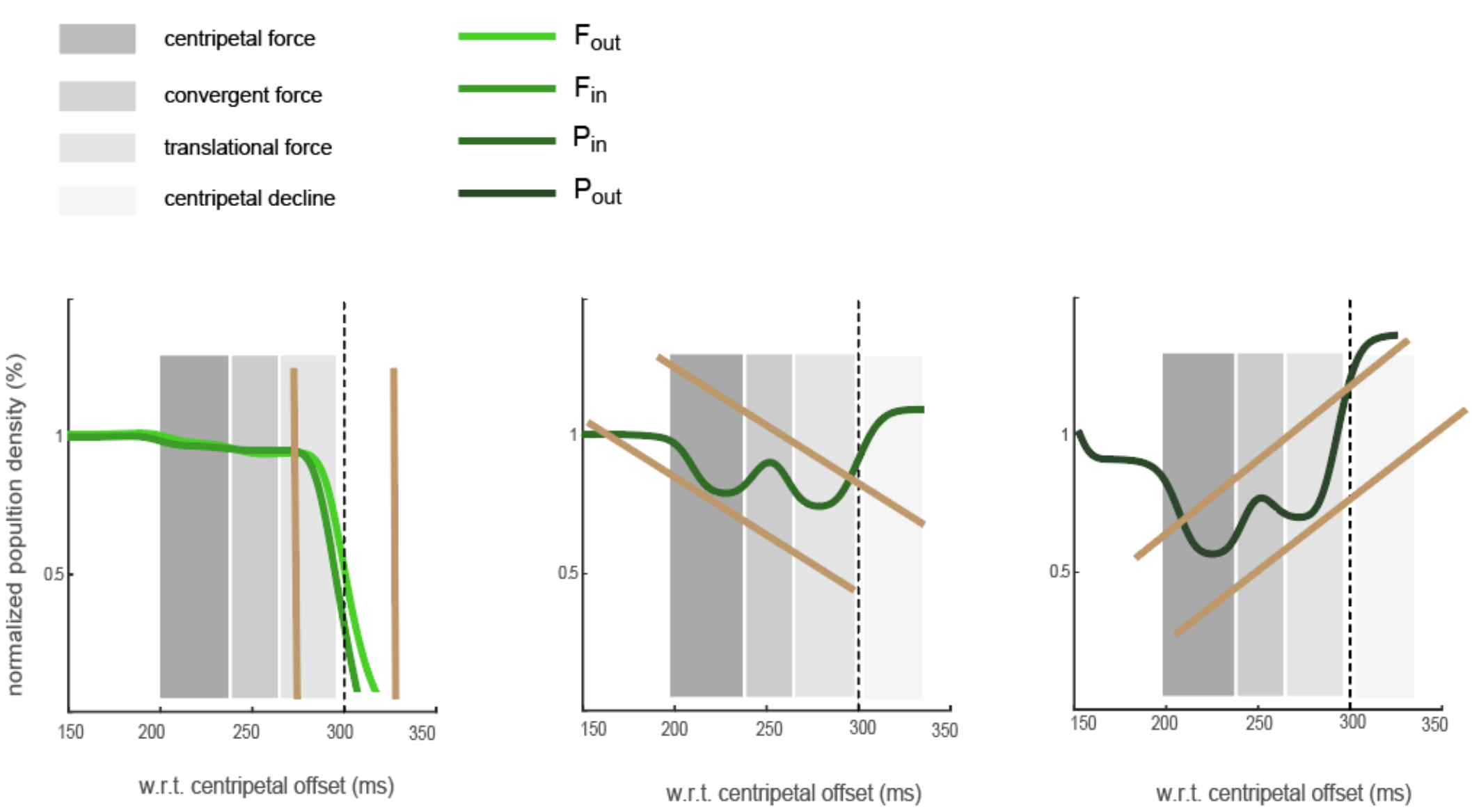

*Figure 4. Simulated changes in density along the radial axis (own representation, Ifedayo-Emmanuel Adeyefa-Olasupo).*

## 4.3. Psychophysical Measurements

In line with predictions from the three-force model, three sets of experiments were conducted with human observers using a cued saccade task. In these experiments, 11 subjects were asked to indicate, via a push button, whether they could detect a faint Gabor probe presented at one of four locations along the radial axis around the time of a SEM. Consistent with simulated predictions, visual sensitivity levels were sustained at $F_{in}$ and $F_{out}$ between 300 and 200 ms before saccade onset, at the expense of the $P_{in}$ and $P_{out}$ extents. These observed responses within this temporal window support the foveated system's ability to maintain high sensitivity within the central 2 degrees of the visual field just before a new target is detected at more eccentric extents of visual space. Notably, this centripetally driven sensitivity has not been previously reported. In the post-saccadic window, an interesting effect emerges, which the simulation also predicts: the extent that is most distal from the saccade target, $F_{out}$, exhibits slightly higher density levels than the $F_{in}$ extent.

Before and after the early pre-saccadic window, during which the convergent force is most influential, a transient increase in visual sensitivity is observed at both the $P_{in}$ and $P_{out}$ extents, with a slightly larger increase at $P_{in}$. Additionally, just before and after the post-saccadic window, $P_{out}$, despite being farther from the central region where resource transfers are expected to be delayed, shows a faster and more pronounced increase in visual sensitivity compared with $P_{in}$. This effect arises just before fixation on the new target location, underscoring the complex dynamics in visual sensitivity distribution across the visual field.

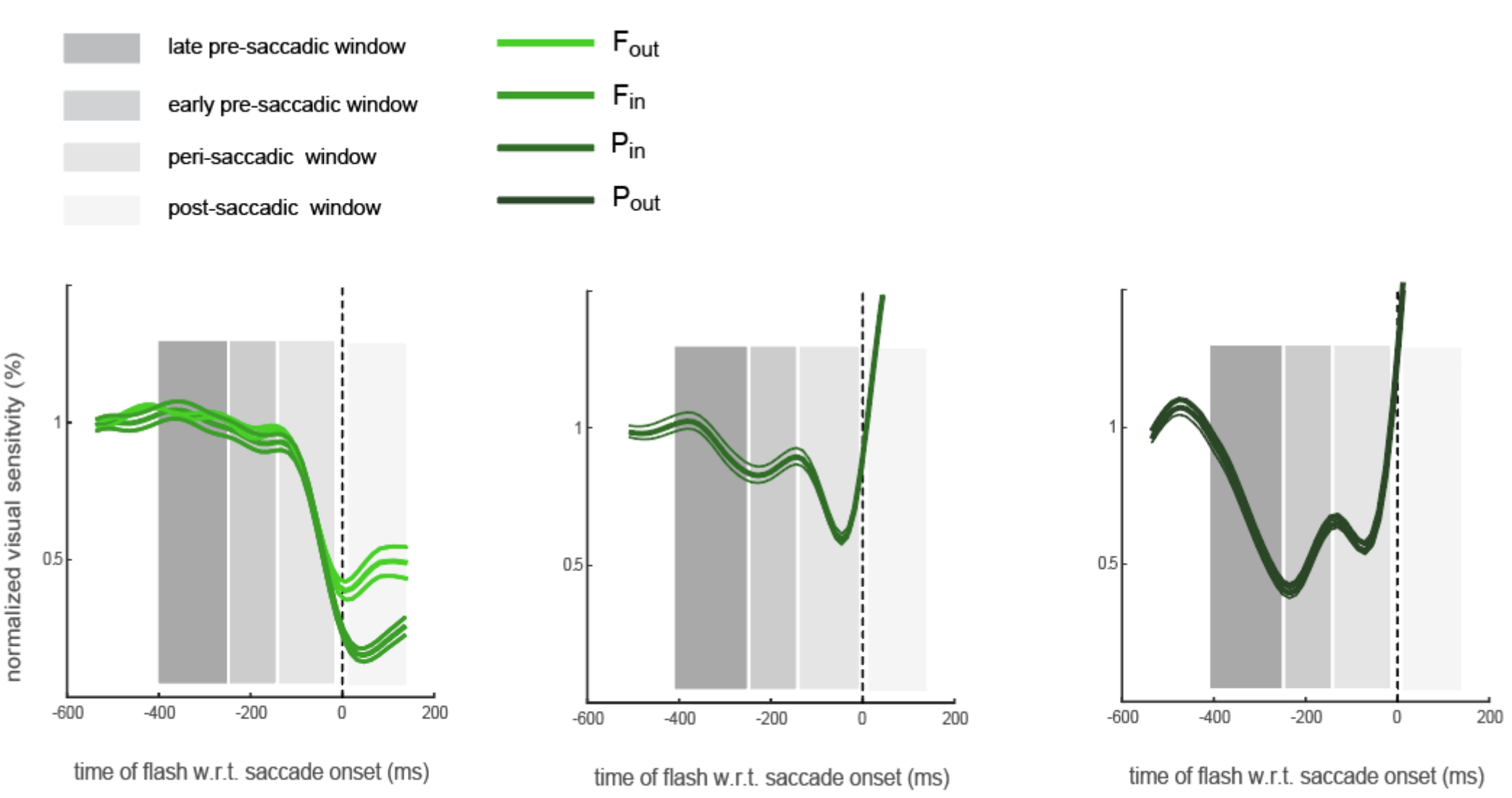

*Figure 5. Experimental changes in sensitivity along the radial axis (own representation, Ifedayo-Emmanuel Adeyefa-Olasupo).*

## 4.4. Paradox I

There are two paradoxes that this general model predicts, which are even more apparent in the empirical results. The first paradox is that the probed location furthest from the saccade target, both before and after the saccade is made, should not experience a higher level of sensitivity than the foveal location that is closer to the target. The second paradox is that the peripheral location furthest from where neural resources are being transferred should not experience a faster and more pronounced increase in visual sensitivity compared with the location closer to this energy source.

The first paradox can be explained by the presence of elastic constraints. Specifically, as the cRF shifts toward the peripheral mass, the cRF located to the left of this mass reaches the boundary of its elastic field and is unable to move further, forcing it to retain its neural energy in the central region. In contrast, the cRF to the right of the central mass can continue shifting in the direction of the saccade target, redistributing its neural resources away from the $F_{in}$ location. This elastic constraint results in the outer foveal location having a slightly higher level of sensitivity than the foveal location closer to the saccade target. This mechanism is essential for biological systems to ensure that peripheral regions are sufficiently sensitised in the presence of potential aversive stimuli, such as a predator.

### 4.5. Paradox II

The second paradox can be explained by considering the tessellation of classical receptive fields and their elastic fields. Specifically, because the size of each cRF scales with eccentricity, when cRFs begin to shift toward the peripheral region, larger cRFs deposit greater amounts of neural resources than smaller cRFs, even when both experience the same magnitude of shift. This size difference, combined with the inverse-distance law, results in larger cRFs producing greater shifts. This in turn leads to the $P_{out}$ location exhibiting a faster and more substantial increase in sensitivity, as it benefits from the larger deposition of neural resources by the larger cRFs just before the eye lands on the target. Notably, since this design principle has only been observed in foveated systems, it would be intriguing to investigate how these effects manifest in other biological systems whose cRF sizes are eccentricity-independent.

# 5. Discussion

The aim of this thesis was to introduce a novel mathematical framework that serves as a general account of active visual processing across different biological systems. Using retinotopic mechanics, I demonstrate that relevant retinotopic brain structures may receive temporally overlapping signals (modelled as forces), causing receptive fields to undergo centripetal, convergent and translational shifts prior to the onset of saccadic eye movements. It is plausible that some cRFs exhibit a preference for one form of remapping over another. However, I show that, across retinotopic brain areas, cRF shifts are actively mediated by their respective elastic fields (εlφ's), and that these shifts follow an inverse-distance law akin to Newton's law of universal gravitation.

While this thesis uncovers the computational relevance of an elastic field, the neurobiological underpinnings of these fields remain unknown. Considering the paradoxes identified in this work, future studies in primates and rodents should aim to replicate these effects and investigate which neurons underlie them. It is likely that, for any animal requiring spatial orientation, elastic fields are fundamental to how the organism orients itself within its environment. In fact, in the absence of elastic fields an animal's ability to perform orientation-mediated tasks, or in extreme cases to survive, would be severely compromised.